\documentclass[aps,prl,reprint,superscriptaddress]{revtex4-2}

\usepackage[T1]{fontenc}
\usepackage[utf8]{inputenc}

\usepackage{graphicx}
\usepackage{amsmath,amssymb}
\usepackage{hyperref}

\begin{document}

\title{Dynamic Lattice Disorder and Collective Dipole Coupling Give Rise to Dicke Physics in Perovskite Quantum Dots}

\author{Priya Nagpal}
\affiliation{Department of Chemistry, McGill University, Montreal, H3A 0B8, Canada}

\author{Patanjali Kambhampati}
\affiliation{Department of Chemistry, McGill University, Montreal, H3A 0B8, Canada}
\email{pat.kambhampati@mcgill.ca}

\begin{abstract}
Halide perovskite quantum dots exhibit cooperative optical phenomena that are absent in conventional semiconductor nanocrystals, including exciton superradiance, superabsorption, and biexciton superradiance within individual dots. Here we develop a microscopic theory that identifies the physical origin of these Dicke effects and establishes how they can be controlled by materials parameters. The central result is that cooperative emission emerges from a competition between collective coupling of optical transition dipoles and lattice-induced disorder, with the balance governed by the Raman-derived phonon spectral density and the excitonic oscillator strength. At elevated temperature, strong Fr\"ohlich coupling and glassy lattice dynamics produce dynamic disorder that suppresses dipole synchronization and yields incoherent emission. Upon cooling, lattice fluctuations freeze and cooperative coherence emerges when the collective coupling exceeds residual static disorder, defining size- and composition-dependent crossover temperatures that we map as phase diagrams. Extending the framework to biexcitons, we show that a confined biexciton constitutes a single correlated charge distribution dressed by a shared lattice configuration, enabling pathway-indistinguishable decay and cooperative radiative enhancement. The theory quantitatively accounts for observed size, composition, and temperature trends in radiative-rate constant ratios and biexciton binding energies, while explaining why full Dicke saturation is not universal. More broadly, the results establish Raman spectral weight and oscillator strength as design parameters for engineering cooperative quantum-optical behavior in quantum materials.
\end{abstract}

\maketitle

Collective light--matter interactions provide a powerful route to enhancing and controlling optical emission in quantum systems\textsuperscript{1-3}. Dicke's prediction\textsuperscript{1} that an ensemble of emitters can synchronize their optical transition dipoles and radiate cooperatively---giving rise to superradiance\textsuperscript{2, 4-10}, and in reverse, superabsorption\textsuperscript{11-15}---revealed that emission rates and light--matter coupling strengths can exceed those of independent emitters by orders of magnitude. These effects underpin a wide range of quantum-optical concepts \textsuperscript{3, 8, 16-17}, from collective energy extraction and directional emission to proposals for quantum information processing and engineered photonic matter. A central challenge is to determine how such cooperative phenomena emerge and remain robust in real quantum materials, where emitters coexist with disorder, phonons, and strong interactions rather than idealized rigid environments.

Halide perovskite quantum dots have recently emerged as an unexpected platform for realizing Dicke physics at the level of a single nanostructure. Superradiance from single excitons in isolated perovskite quantum dots (QD) was first reported by Kambhampati and coworkers in 2023 \textsuperscript{10} and independently by Kovalenko and collaborators in 2024 \textsuperscript{5} on single QD, while superabsorption was subsequently demonstrated by Kovalenko and coworkers \textsuperscript{11}. These observations establish that collective optical coherence can arise within a soft, dynamically disordered, and strongly anharmonic lattice \textsuperscript{18-20}---conditions that are generally expected to suppress cooperative emission. \emph{Ab initio} molecular dynamics simulations were first applied to this problem by Prezhdo and Kambhampati\textsuperscript{10} and later extended by Kovalenko and collaborators\textsuperscript{5}, revealing large-amplitude, ultrafast lattice fluctuations characteristic of a glassy solid\textsuperscript{21-23}. In parallel, picosecond non-Condon photoluminescence dynamics were observed at room temperature and shown to vanish entirely at cryogenic temperature, directly demonstrating a crossover from dynamically fluctuating optical coupling to a frozen-lattice regime\textsuperscript{24}.

Despite these advances, a microscopic understanding of cooperative radiative phenomena in perovskite quantum dots remains incomplete. Existing theoretical approaches emphasize either incoherent exciton--phonon scattering, spectral diffusion, and transport, or phenomenological descriptions of superradiant scaling developed for rigid semiconductor nanocrystals. Such frameworks do not capture the defining physical environment of halide perovskites: a lattice with large Raman spectral weight extending to low energies, strong Fr\"ohlich coupling, substantial dynamic disorder at elevated temperature, and multiexciton Coulomb correlations evolving on comparable energy scales. This places perovskite quantum dots far outside the theoretical regime established for conventional nanocrystals such as CdSe, where weak lattice coupling and narrow phonon spectra localize oscillator strength and suppress Dicke states even at low temperature. What has been missing is a theory that translates lattice response functions into experimentally accessible parameters governing collective optical coherence.

Here we show that Dicke physics in perovskite quantum dots emerges from a competition between collective coupling of optical transition dipoles and lattice-induced disorder, with the balance controlled by the Raman-derived phonon spectral density, oscillator strength, dot size, and composition. By developing a microscopic density-matrix and collective-state framework, we identify how dynamic lattice disorder suppresses cooperative emission at elevated temperature, while frozen lattice configurations at low temperature enable exciton superradiance, superabsorption, and biexciton superradiance. This approach yields quantitative phase diagrams that connect radiative-rate ratios and biexciton binding energies directly to materials parameters, providing a unified explanation of existing experiments and predictive design rules for engineering cooperative quantum-optical behavior in quantum materials.

Figure~1 introduces the physical setting and organizing principles underlying cooperative emission in perovskite quantum dots. Panels (a) and (b) contrast two limiting regimes for an ensemble of optically active unit cells within a single dot. At elevated temperature [Fig.~1(a)], strong lattice fluctuations randomize the phases of individual optical transition dipoles, suppressing coherent addition and yielding effectively independent emission. At low temperature [Fig.~1(b)], lattice disorder is reduced and collective dipole--dipole coupling can synchronize phases across many unit cells, producing a macroscopic bright state with an enhanced radiative rate constant. The crossover between these regimes is not a thermodynamic phase transition, but a coherence crossover governed by the competition between collective coupling and disorder.

Panels (c) and (d) highlight why halide perovskites uniquely access this regime. The perovskite lattice combines large unit-cell transition dipoles with a soft, glassy vibrational environment often described as a ``phonon glass--electron crystal'' \textsuperscript{21-23, 25-28}. Raman spectroscopy reveals a broad spectral density extending to low frequencies with large Huang--Rhys factors, reflecting strong Fr\"ohlich coupling and substantial dynamic disorder\textsuperscript{25}. In contrast, conventional nanocrystals such as CdSe exhibit weak electron--phonon coupling and narrow, well-resolved phonon modes\textsuperscript{25, 29}. As a result, perovskite quantum dots naturally sit near the boundary between incoherent and cooperative emission, whereas rigid nanocrystals remain deep in the incoherent regime. Panels (e) and (f) summarize the experimental consequences: strong enhancement of the exciton radiative rate constant at low temperature and an anomalous size dependence that is absent at room temperature.

These observations motivate a microscopic theory in which cooperative emission emerges from the balance between collective dipole coupling and lattice-induced disorder. Details of the theory can be found in the Supplementary Information (SI). We begin by considering a single bright exciton confined within a quantum dot and coupled to lattice vibrations. The minimal Hamiltonian is
\begin{equation}
H = \hslash\omega_{X} \, |X\rangle\langle X| + \sum_{q}\hslash\omega_{q}\, b_{q}^{\dagger}b_{q} + \sum_{q} g_{q}(b_{q} + b_{q}^{\dagger})\, |X\rangle\langle X| ,
\end{equation}
where $|X\rangle$ denotes the optically allowed exciton state, $b_{q}$ are phonon operators, and $g_{q}$ describes exciton--phonon coupling. All lattice effects enter through the phonon spectral density
\begin{equation}
J(\omega) = \sum_{q} |g_{q}|^{2}\,\delta(\omega - \omega_{q}),
\end{equation}
which is directly accessible experimentally through Raman spectroscopy. For perovskites, $J(\omega)$ is broad and weighted toward low frequencies, whereas for CdSe it is narrow and weak.

The phonon bath induces fluctuations of the exciton transition energy and, equivalently, fluctuations of the optical transition dipole. These fluctuations generate a disorder scale
\begin{equation}
\sigma^{2}(T)=\int_{0}^{\infty} d\omega\, J(\omega)\,\coth\!\left(\frac{\hslash\omega}{2k_{B}T}\right),
\end{equation}
which is large and temperature dependent in perovskites. At room temperature, dynamic lattice disorder dominates, leading to rapid dephasing and incoherent emission. Upon cooling, phonon occupation decreases and dynamic disorder freezes out, leaving a residual static disorder floor determined by the low-frequency spectral weight.

Collective emission arises when this disorder scale becomes smaller than the collective dipole--dipole coupling between unit cells. Comparison of the transient absorption bleaching response of anionic CsPbBr$_3$ perovskite QD relative to CdSe QD of the same diameter shows that perovskite has 10 times greater absorption cross section, hence transition dipole moments\textsuperscript{30-31}. The radiative rate constant of the exciton can then be written as $k_{r}^{X}(T,D)=N_{\mathrm{coh}}(T,D)\,k_{0}^{X}$, where $k_{0}^{X}$ is the single-unit-cell radiative rate constant and $N_{\mathrm{coh}}$ is the number of unit cells whose transition dipoles are phase locked. The coherence number grows with dot size and decreases with disorder, reflecting the competition between collective coupling and lattice fluctuations.

Experimentally, this competition is conveniently captured through the ratio
\begin{equation}
R_{X}(T,D)=\frac{k_{r}^{X}(T,D)}{k_{r}^{X}(300\,K,D)}=\exp\!\left[A_{X}(D)\left(\frac{1}{T}-\frac{1}{300}\right)\right],
\end{equation}
where the energy scale $A_{X}(D)$ encodes the net coherence energy set by collective coupling minus residual disorder. This form directly explains the strong enhancement of exciton radiative rate constants at low temperature, the absence of enhancement at room temperature, and the anomalous size dependence observed in Fig.~1(f). The crossover temperature $T_{c}$, defined by a fixed enhancement threshold, increases with dot size and depends systematically on composition through the Raman spectral density and oscillator strength.

Figure~2 connects the microscopic theory to the experimental phenomenology of exciton superradiance. Panel (a) shows the size dependence of the energy scale $A_{X}(D)$ extracted from temperature-dependent radiative-rate ratios for different compositions. Within the theory, $A_{X}$ represents the net coherence energy that governs the growth of the excitonic coherence volume upon cooling. Microscopically, it reflects the competition between collective dipole--dipole coupling, which increases with dot size and oscillator strength, and lattice-induced disorder set by the Raman spectral density. The observed increase of $A_{X}$ with diameter directly indicates that larger dots support stronger collective coupling, allowing coherence to extend across a greater number of unit cells once dynamic disorder is suppressed. The systematic composition dependence of $A_{X}$ further confirms that lattice softness and phonon spectral weight---rather than electronic confinement alone---control the approach to cooperative emission.

Panel (b) presents the ratio $R_{X}=\frac{k_{r}^{X}(10\,K)}{k_{r}^{X}(300\,K)}$, as a function of diameter. Results are shown from calculations and from experiment by Kambhampati\textsuperscript{10} and Kovalenko\textsuperscript{5}. There is excellent correspondence between experiment and theory. The strong superlinear increase of $R_{X}$ with size is a hallmark of cooperative behavior: as the coherence number $N_{\mathrm{coh}}$ grows, the radiative rate constant increases proportionally. In contrast, at room temperature the same dots exhibit little or no size dependence, reflecting the dominance of dynamic lattice disorder that limits coherence to a small number of unit cells. This striking contrast between low- and high-temperature behavior is a direct manifestation of the disorder--coupling competition captured by the theory. Importantly, the enhancement is expressed in terms of radiative rate constants rather than emission intensities, demonstrating that the effect arises from intrinsic changes in light--matter coupling rather than population dynamics.

Panels (c) and (d) summarize these trends in the form of phase diagrams. Panel (c) shows the temperature--diameter phase boundary for CsPbBr$_3$, defined by a fixed enhancement threshold. The increase of the crossover temperature $T_{c}$ with dot size follows naturally from the increase in collective coupling energy with coherence volume: larger dots require less suppression of lattice disorder to enter the cooperative regime. Panel (d) extends this analysis to composition at fixed diameter, revealing a clear ordering of $T_{c}$ that tracks lattice spectral weight. Compositions with softer lattices and larger low-frequency Raman weight exhibit lower $T_{c}$, while stiffer lattices with reduced disorder support cooperative emission at higher temperatures. These phase diagrams demonstrate that exciton superradiance in perovskite quantum dots is not accidental, but a tunable consequence of materials parameters that can be rationally controlled.

Taken together, Fig.~2 establishes that exciton superradiance\textsuperscript{4-5, 10} is determined by three interrelated factors: the strength of the optical transition dipole, the size-dependent coherence volume, and the magnitude of lattice-induced disorder encoded in the phonon spectral density. Superradiance emerges when collective coupling exceeds disorder, and its size and composition dependence directly reflects how close a given material lies to this balance. This framework explains why cooperative emission is absent in rigid nanocrystals such as CdSe and why perovskite quantum dots uniquely access a regime where exciton coherence can be tuned across a wide temperature range. In the following, we show how the same physical competition governs biexciton binding and cooperative biexciton emission, with additional constraints imposed by many-body structure.

We now turn to biexcitons\textsuperscript{30, 32}, where many-body correlations introduce additional structure but the same competition between collective coupling and lattice disorder continues to govern cooperative emission. A biexciton confined within a quantum dot is a correlated four-particle state composed of two electrons and two holes occupying the same spatial volume. Crucially, it constitutes a single many-body charge distribution, rather than two independent excitons. As a result, the lattice responds to the biexciton collectively, coupling to the total charge density rather than to spatially distinct centers. This distinction has important consequences for both the biexciton binding energy and the nature of its radiative decay.

The biexciton binding energy can be written schematically as $\Delta_{XX}=\Delta_{\mathrm{Coul}}+\Delta_{\mathrm{lat}}$ where $\Delta_{\mathrm{Coul}}$ arises from direct Coulomb interactions between quasiparticles and $\Delta_{\mathrm{lat}}$ reflects lattice-mediated stabilization. In rigid nanocrystals such as CdSe\textsuperscript{32-35}, weak Fr\"ohlich coupling and a narrow phonon spectrum render the lattice contribution negligible, so the binding energy is dominated by Coulomb interactions and decreases rapidly with increasing dot size. In halide perovskite quantum dots, by contrast, strong lattice polarizability and broad low-frequency Raman spectral weight produce a substantial lattice contribution at elevated temperature, yielding comparatively large biexciton binding energies with a much weaker size dependence.

These distinctions are evident in Fig.~3. The measured biexciton binding energies, plotted as a function of normalized diameter, show a steep size dependence in CdSe that is well described by Coulomb scaling, whereas CsPbBr$_3$ exhibits a flatter dependence with significantly larger absolute binding energies. Isolating the lattice contribution reveals that it increases with lattice softness and decreases upon cooling. This temperature dependence reflects the same lattice dynamics that control exciton coherence: at room temperature, dynamic lattice fluctuations stabilize the biexciton through polaronic dressing, while at low temperature phonon occupation collapses and dynamic lattice contributions freeze out.

Importantly, the reduction of the lattice contribution at low temperature does not imply that the biexciton becomes decoupled from the lattice. Rather, the biexciton freezes into a single static lattice configuration shared across the entire charge distribution. This picture is fundamentally different from a bipolaronic scenario, in which independent lattice distortions accompany separate charges. In the single-polaron picture appropriate for confined biexcitons, the lattice does not encode which-path information associated with different radiative decay channels. This absence of distinguishability is a prerequisite for cooperative biexciton emission.

These considerations naturally connect biexciton binding to biexciton superradiance. Radiative decay of a biexciton proceeds through two indistinguishable pathways corresponding to recombination of either electron--hole pair. When the lattice environment is shared and does not distinguish between these pathways, the corresponding transition amplitudes interfere constructively. The biexciton radiative rate constant can then approach twice that of the exciton, defining the Dicke limit. We quantify this effect through the ratio,
At elevated temperature, dynamic lattice disorder redistributes population across multiple biexciton substates and encodes which-path information, yielding . Upon cooling, suppression of dynamic disorder funnels population into a small number of bright states and removes pathway distinguishability, allowing cooperative enhancement to emerge.

The observed temperature, size, and composition dependence of  follows directly from this picture. Larger dots support a larger coherence volume and therefore stronger collective coupling, while compositions with softer lattices and greater low-frequency Raman weight retain larger residual disorder. As a result, full saturation at the Dicke limit is achieved only for selected sizes and compositions, while other systems remain subsaturated even at cryogenic temperature. This behavior highlights that lowering temperature alone is not sufficient to guarantee biexciton superradiance; suppression of residual static disorder and collapse of the biexciton manifold into a single bright state are also required.

Together with the exciton results, these findings establish a unified microscopic description of Dicke physics in perovskite quantum dots\textsuperscript{3, 5, 10-11}. Exciton superradiance, superabsorption, and biexciton superradiance arise from the same competition between collective dipole coupling and lattice-induced disorder, with crossover temperatures, enhancement factors, and saturation limits determined by dot size, composition, oscillator strength, and the Raman-derived phonon spectral density. The theory explains why such cooperative effects are absent in rigid nanocrystals such as CdSe and why perovskites uniquely access a regime where they can be tuned.

More broadly, the framework provides concrete design principles for engineering cooperative quantum-optical behavior in quantum materials. Strong unit-cell transition dipoles, broad low-frequency lattice spectral weight, and controlled suppression of static disorder emerge as key requirements for realizing Dicke phases. Within the family of halide perovskites, these criteria can be addressed through composition, size, and dielectric environment. Beyond perovskites, the results suggest a general strategy for designing soft quantum solids in which lattice dynamics and collective light--matter coupling are balanced to produce robust cooperative emission.

\medskip
Supplementary Information.

Details of the theory and fitting are provided in the SI.

\begin{acknowledgments}
Financial support from NSERC and McGill University is acknowledged.
\end{acknowledgments}

% =========================
% References
% =========================

% =========================
% Figures (PNG files)
% =========================

\begin{figure}[t]
\centering
\includegraphics[width=\linewidth]{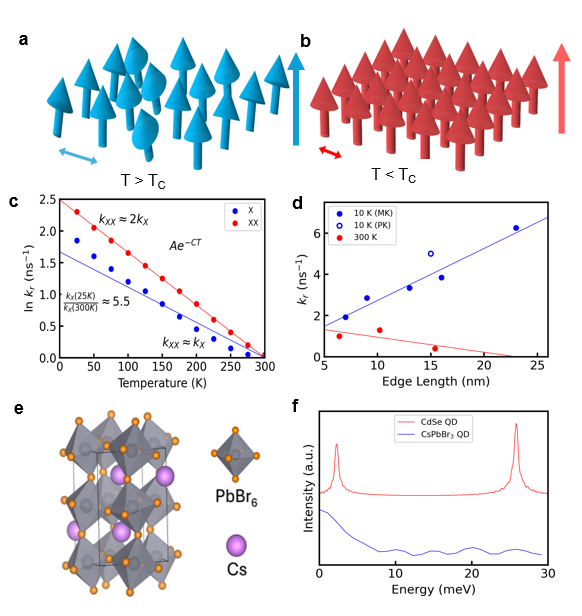}
\caption{Figure 1 | Emergence of Dicke physics in perovskite quantum dots and its lattice origin. (a) Schematic illustration of an ensemble of unit-cell transition dipoles at high temperature (T$>$T$_c$), where strong lattice-induced disorder randomizes dipole phases and suppresses cooperative emission. (b) At low temperature (T$<$T$_c$), suppression of dynamic disorder allows dipole--dipole coupling to synchronize the unit-cell dipoles into a macroscopic polarization, giving rise to Dicke superradiance (and, in reverse, superabsorption). (c) Experimental observation of superradiance from both excitons (X) and biexcitons (XX) in a single 15-nm CsPbBr$_3$ quantum dot, first reported by Kambhampati and coworkers (2023), showing a strong enhancement of the radiative rate constant upon cooling. (d) Single-dot measurements of the size dependence of exciton superradiance in CsPbBr$_3$ quantum dots reported by Kovalenko and collaborators (2024), demonstrating a pronounced low-temperature enhancement that increases with dot diameter, in contrast to the weak and non-systematic behavior at room temperature. (e) Structural motif of the halide perovskite lattice, highlighting its soft, dynamically disordered, and defect-tolerant nature, which enables strong coupling between electronic excitations and lattice degrees of freedom. (f) Raman spectra comparing a perovskite quantum dot and a CdSe quantum dot, illustrating the broad, glassy low-frequency spectral weight and strong exciton--phonon coupling (S$\sim$1) in perovskites, in contrast to the weak, discrete phonon modes (S$\sim$0.1) of CdSe. Together, these panels motivate a microscopic picture in which the balance between collective dipole coupling and lattice-induced disorder governs the emergence of Dicke physics in perovskite quantum dots.}
\end{figure}

\begin{figure}[t]
\centering
\includegraphics[width=\linewidth]{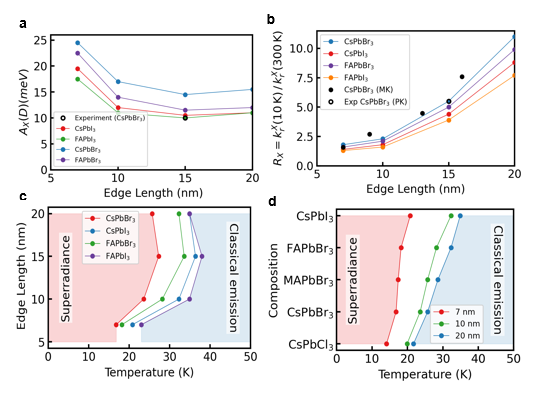}
\caption{Figure 2 | Exciton superradiance in perovskite quantum dots: size, composition, and phase behavior. (a) Size dependence of the energy parameter that controls the temperature evolution of the exciton radiative rate constant. This parameter is extracted from the exponential dependence of the radiative-rate constant enhancement on inverse temperature and represents the effective coherence energy arising from the competition between collective dipole coupling and lattice-induced disorder. Its systematic variation with dot diameter and halide composition reflects changes in coherence volume and lattice spectral weight. (b) Corresponding enhancement ratio of the exciton radiative rate constant at 10 K relative to 300 K, showing a strong, superlinear increase with diameter and a clear composition dependence, indicative of the rapid growth of cooperative coherence at low temperature. (c) Temperature--diameter phase diagram separating incoherent and Dicke-superradiant regimes, constructed using a fixed enhancement threshold. Larger dots enter the superradiant regime at higher temperature, reflecting stronger collective coupling. (d) Temperature--composition phase diagram for selected diameters, illustrating how lattice softness and phonon spectral weight tune the crossover temperature between incoherent and superradiant phases. Together, these panels demonstrate that exciton superradiance in perovskite quantum dots is governed by a balance between collective dipole coupling and lattice-induced disorder, and can be systematically tuned by size and composition.}
\end{figure}

\begin{figure}[t]
\centering
\includegraphics[width=\linewidth]{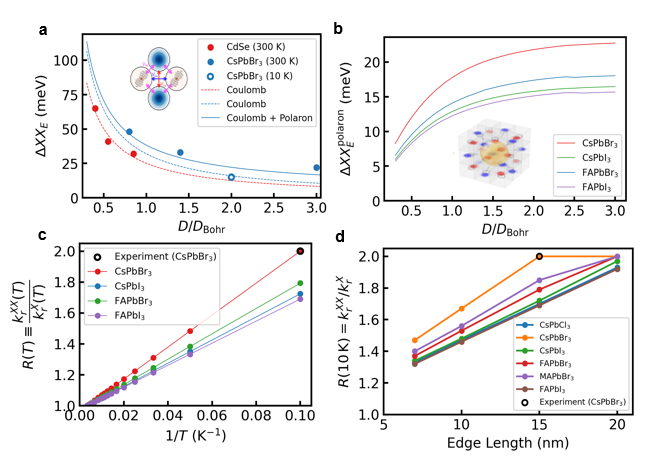}
\caption{Figure 3 | Biexciton Dicke physics in perovskite quantum dots: binding, lattice dressing, and cooperative emission. (a) Size-dependent biexciton binding energies extracted from photoluminescence for CdSe and CsPbBr$_3$ quantum dots at 300 K, plotted versus normalized diameter (D/a$_B$). The curves compare Coulomb-only behavior with Coulomb plus lattice stabilization. For perovskites, removing lattice stabilization yields the limiting behavior appropriate to the low-temperature ($\approx$10 K) regime, where the apparent biexciton binding energy is reduced by approximately a factor of two relative to 300 K. (b) Lattice (polaronic) contribution to the biexciton binding energy versus (D/a$_B$) for representative perovskite compositions. The increase and subsequent saturation with size reflects the formation of a shared, spatially coherent lattice polarization field dressing the biexciton, consistent with a single-polaron (not bipolaron) picture and enabling pathway indistinguishability required for cooperative biexciton emission. (c) Predicted temperature dependence of biexciton superradiance quantified by the ratio of biexciton to exciton radiative rate constants, plotted versus inverse temperature for a fixed dot size and multiple compositions. The approximately linear dependence on (1/T) reflects progressive suppression of lattice-induced dephasing and funneling into bright biexciton states upon cooling. (d) Low-temperature (10 K) size dependence of the biexciton-to-exciton radiative-rate constant ratio for multiple compositions, illustrating how closely the system approaches the Dicke limit of 2 and how incomplete saturation depends on size and lattice/compositional disorder.}
\end{figure}

\end{document}